\documentclass[aps,prd,onecolumn,amssymb]{revtex4}
\usepackage{graphicx,bm,color}
\usepackage{amsmath}
\usepackage{amssymb}
\usepackage{amsfonts}
\usepackage{epsfig}
\newcommand{\be}{\begin{equation}}
\newcommand{\ee}{\end{equation}}
\newcommand{\bea}{\begin{eqnarray}}
\newcommand{\eea}{\end{eqnarray}}
\newcommand{\beaa}{\begin{eqnarray*}}
\newcommand{\eeaa}{\end{eqnarray*}}

\allowdisplaybreaks[4]

\begin{document}

\title{ Gravitational wave pulse and memory effects for hairy Kiselev black hole and its analogy with Bondi-Sachs formalism}
\author{H. Hadi$^1$}\email{hamedhadi1388@gmail.com}
\author{Amin Rezaei Akbarieh$^1$}\email{am.rezaei@tabrizu.ac.ir}
\author{David F. Mota$^2$}\email{d.f.mota@astro.uio.no}
\affiliation{$^1$Faculty of Physics, University of Tabriz
Tabriz 51666-16471, Iran\\ $^2$Institute of Theoretical Astrophysics, University of Oslo, Sem S$\backslash$ae$\{$lands$\}$ vei 13, 0371 Oslo, Norway}

\begin{abstract}
	The investigation of non-vacuum cosmological backgrounds containing black holes is greatly enhanced by the Kiselev solution. This solution plays a crucial role in understanding the properties of the background and its relationship with the features of the black hole. Consequently, the gravitational memory effects at large distances from the black hole offer a valuable means of obtaining information about the surrounding field parameter $N$ and parameters related to the hair of the hairy Kiselev Black hole. 
This paper investigates the gravitational memory effects in the context of the Kiselev solution through two distinct approaches. At first, the gravitational memory effect at null infinity is explored by utilizing the Bondi-Sachs formalism by introducing a gravitational wave (GW) pulse to the solution. The resulting Bondi mass is then analyzed to gain further insight. Therefore, the Kiselev solution is being examined to determine the variations in Bondi mass caused by the pulse of GWs. The study of changes in Bondi mass is motivated by the fact that it is dynamic and time-dependent, and it measures mass on an asymptotically null slice or the densities of energy on celestial spheres. In the second approach, the investigation of displacement and velocity memory effects is undertaken in relation to the deviation of two neighboring geodesics and the deviation of their derivative influenced by surrounding field parameter $N$ and the hair of hairy Kiselev black hole. This analysis is conducted within the context of a gravitational wave pulse present in the background of a hairy Kiselev black hole surrounded by a field parameter N.
\end{abstract}
\maketitle

\section{Introduction}
The detection of gravitational waves (GWs) has been a significant advancement in the field of gravitational physics over the past decade. Two major observational projects have been focused on detecting GWs from binary black holes and neutron star merger events. \cite{binary1,binary2}. The study of the shadow of central objects with supermassive compactness is a significant experimental endeavor in this domain, as evidenced by various sources \cite{massive1,massive2,massive3,massive4,massive5,massive6,massive7}. The behavior of gravity in strong fields is a critical factor in these observations, which serve as a crucial test for the principles of general relativity and provide insights into the characteristics of compact objects.  \cite{compact1,compact2,compact3,compact4,compact5,compact6,compact7,compact8,compact9}.

The present study aims to investigate certain characteristics of the Kiselev black hole solution while considering the presence of GWs in the background of the black hole. The reason behind examining this particular solution stems from the fact that, within the domain of astrophysics, black holes do not exist in isolation but rather coexist within non-vacuum environments. A portion of scientific inquiry has been dedicated to exploring cosmic backgrounds' immediate and localized impacts on the established black hole solutions. For example, it has been demonstrated that \cite{babichev} within a universe undergoing expansion due to the presence of a phantom scalar field, the mass of a black hole diminishes as a consequence of the accumulation of particles from the phantom field into the central black hole. Nevertheless, it is important to acknowledge that this effect is of a global nature. To investigate the localized alterations in the spacetime geometry in the vicinity of the central black hole, it is necessary to consider a modified metric that incorporates the surrounding spacetime. In this regard, Kiselev has provided an analytical solution to the Einstein field equations, which describes a static spherically symmetric scenario \cite{kiselevx}. The generalization of the Schwarzschild black hole to a non-vacuum background is a solution that exhibits an effective equation of state parameters for the surrounding field of the black hole. This allows for a wide range of possibilities, including quintessence, cosmological constant, radiation, and dust-like fields. Furthermore, the dynamical Vaidya-type solutions have also been generalized based on this initial solution \cite{22y,23y,24y}. These generalizations are well-founded due to the non-isolated nature of real-world black holes and their existence in non-vacuum backgrounds. Black hole solutions coupled with matter fields, such as the Kiselev solution, hold particular relevance in the study of astrophysical black holes with distortions \cite{25y,26y,27y,28y}. Additionally, they play a significant role in investigating the no-hair theorem \cite{29y,30y,31y,32y}. This theorem asserts that a black hole can be fully described by three charges, namely mass ($M$), electric charge ($Q$), and angular momentum ($a$). However, it relies on the crucial assumption that the black hole is isolated, meaning that the spacetime is asymptotically flat and devoid of other sources.

Therefore, studying the Kiselev black hole as a representative example of a genuine black hole in the natural world holds significant value. Identifying and determining its field parameter $N$ through the analysis of gravitational memory effects can prove beneficial. In this particular context, we explore specific attributes of the Kiselev black hole solution, taking into account the existence of gravitational waves in its background. This analysis focuses on the Bondi-Sachs formalism  \cite{bondii,ref1}, which is a novel approach to studying gravitational waves in the context of general relativity. This formalism, as proposed in \cite{bondi1}, utilizes outgoing null rays that follow the path of the waves in axisymmetric spacetimes. The primary objective of this research is to detect gravitational memory effects associated with the Kiselev black hole solution. The extension of this formalism to spacetime that is not axisymmetric and the determination of the asymptotic symmetries as one approaches infinity along the outgoing null hypersurfaces has been accomplished in the work by Bondi-Sachs \cite{bondi1,bondi2}.
The investigation conducted in \cite{ref2} focused on the limitations surrounding the three potential sources of gravitational memory effect within the linearized Bondi-Sachs framework. The findings revealed that the sole known origin of B-mode gravitational memory is of primordial nature. In the linearized theory, this corresponds to a uniform wave entering from past null infinity. A comprehensive study in \cite{ref3} delved into the relationship between gravitational memory effects and the Bondi-Metzner-Sachs symmetries of asymptotically flat spacetimes within the scalar-tensor theory. Within this study, the solutions to the equations of motion near the future null infinity were obtained using the generalized Bondi-Sachs coordinates, while adhering to a suitable determinant condition. The extension of Bondi-Sachs form to alternative modified gravity theories like Brans-Dicke theory was examined in \cite{ref4}. The research investigated the correlation between memory effects, symmetries, and conserved quantities within Brans-Dicke theory. Furthermore, the field equations were calculated in Bondi coordinates, and a specific set of boundary conditions were established to represent asymptotically flat solutions within this framework. Subsequently, the determination of the asymptotic symmetry group for these spacetimes revealed that it aligns with the Bondi-Metzner-Sachs group in general relativity. In the publication \cite{ref5}, the authors revisit the metric derivation in Newman-Unti and Bondi gauges, specifically focusing on the complete quadrupole-quadrupole interactions. They proceed to rederive the displacement memory effect and provide expressions for all relevant Bondi aspects and dressed Bondi aspects, which are crucial for studying both the leading and subleading memory effects. Additionally, they successfully determine the Newman-Penrose charges, the BMS charges, and the celestial charges of second and third order. These charges are defined based on the known second order and novel third order dressed Bondi aspects, considering interactions involving mass monopole-quadrupole and quadrupole-quadrupole. The investigation of the gravitational wave (GW) memory is conducted within the framework of a specific category of braneworld wormholes, which possess a unique characteristic of not requiring any exotic matter fields for their traversability as mentioned in \cite{ref6}. The memory effect, which signifies the imprint of the presence of extra dimensions and the wormhole nature of the spacetime geometry, is discussed in \cite{ref6}. Furthermore, it has been suggested in \cite{ref7,ref8} that the charges associated with the internal Lorentz symmetries of general relativity, along with the inclusion of higher derivative boundary terms in the action, can capture observable gravitational wave effects.

The Bondi-Sachs formalism employs six metric quantities to characterize the general spacetime. This formalism can be employed to investigate the memory effects in General Relativity  or an altered theory of it. 
The trace of gravitational waves (GW) in the background of a metric, which satisfies Einstein's equations, is known as the gravitational memory effect. This phenomenon, which has been discovered  a long time ago \cite{m1,m2,m3,m4}, arises due to the propagation of energy flux towards the future null infinity, where the spacetime is asymptotically flat \cite{x1}. The region in which this effect occurs is characterized by an asymptotic symmetry group known as the Bondi-Metzner-Sachs (BMS) group \cite{bondii,bondi1,bondi2}. 

Practically, the investigation of the infrared structure of gravity holds significant importance for physicists, with the memory effect being a key aspect of interest\cite{infrared1,infrared2,infrared3}. Another intriguing area of study concerning the gravitational memory effect arises from the detection of GWs, as referenced in \cite{GW1,GW2,GW3,GW4,GW5,GW6,GW7,GW8}. 

In the current context, our objective is to ascertain the trace of memory effects pertaining to Kiselev black hole solutions, which serve as a compact central object for examination. To achieve this, we employ the Bondi-Sachs formalism and endeavor to identify the imprint of memory effects associated with the parameters of the black hole from null infinity. Subsequently, we shall investigate the gravitational memory effects by analyzing the deviation of two neighboring geodesics within this solution in the presence of a GW pulse.
The paper is structured as follows. We bring  a brief review of Kiselev black hole solution and Bondi-Sachs formalism in sections (II) and (III), respectively. Section (IV) provides a comprehensive review of the hairy Kiselev solution, followed by an examination of the memory effects and the alteration in Bondi mass through the application of the Bondi-Sachs formalism. Section (V) delves into the study of memory effects, specifically focusing on the deviation of two neighboring geodesics and their derivatives as displacement and velocity memory effects. Finally, the paper concludes with a dedicated section for summarizing the findings.

 \section{Kiselev black hole }
 
The present section delves into the formulation of the Bondi-Sachs formalism for a hairy Kiselev solution, in a manner that is consistent with \cite{771,6,7}. The gravitational decoupling method is employed in the construction of the hairy Schwarzschild solution, as described in \cite{hairy}. In this context, we provide a brief overview of the solution and subsequently, in the next sections, incorporate the GW components, before analyzing it within the framework of the Bondi-Sachs formalism.
 
 In the gravitational decoupling method, the energy-momentum tensor is written as
 \begin{equation}
 	\bar{T}_{ik}=T_{ik}+\mathcal{T}_{ik},
 \end{equation}
 where $T_{ik}$ represents the energy-momentum tensor in Einstein equation $G_{ik}=8\pi G T_{ik}$ and the solution for it is supposed to be known. The extra matter source $\mathcal{T}_{ik}$ causes an extra geometric deformation as 
$\mathcal{G}_{ik}=\alpha \mathcal{T}_{ik}$. Although the Einstein equation  possesses a non-linear solution, it is possible to express it in a direct superposition of two solutions, as follows: 
 \begin{equation}\label{effectiveGR}
 	\bar{\mathcal{G}}_{ik}=8\pi G T_{ik}+\alpha\mathcal{T}_{ik}.
 \end{equation} 
 Now, assuming the solution of Einstein's equation  $G_{ik}=8\pi G T_{ik}$ , we consider the provided seed metric as
\begin{equation}\label{seed}
 	ds^{2}=-e^{\nu(r)}dt^{2}+e^{\lambda(r)}dr^{2}+r^{2}d\Omega^{2}.
 \end{equation}
Here $d\Omega^{2}$ is the metric on unit two-sphere and the functions $\nu(r)$ and $\lambda(r)$ solely depend on the $r$ coordinate and are assumed to be known. The geometrical deformation of $(\ref{seed})$ by the extra energy-momentum tensor with new functions $\zeta=\zeta(r)$ and $\eta=\eta(r)$ is given by
 \begin{equation}\label{new}
 	e^{\nu(r)}\rightarrow e^{\nu(r)+\alpha \zeta(r)}~~~~~~~~~,~~~~~~~e^{\lambda(r)}\rightarrow e^{\lambda(r)}+\alpha \eta(r),
 \end{equation}
where $\alpha$ is coupling constant and $\eta$ ,$\zeta$ are because of geometrical deformation. The metric components undergo deformations due to the influence of the new matter source $\mathcal{T}_{ik}$. When $\zeta=0$, only the $g_{11}$ component is impacted, while the $g_{00}$ component remains unchanged. This phenomenon is referred to as the minimal geometrical deformation. However, this approach has certain limitations. For example, it fails to explain the existence of a stable black hole with a well-defined event horizon \cite{yag}. A more comprehensive framework is provided by the extended gravitational decoupling, which permits deformations in both the $g_{00}$ and $g_{11}$ components. This methodology is only applicable to the vacuum seed solutions of the Einstein equations, as it violates the Bianchi identities in the presence of a matter source, except in specific cases. One such case is the Vaidya solution, which can be deformed without losing the property of extended gravitational decoupling. However, this property does not hold for the generalized Vaidya solution, as there is an exchange of energy between two matter sources \cite{yag2}. The influence of a primary hair on the thermodynamics of a black hole (\ref{seed}) has been examined in \cite{silva1}.
The Einstein equation for the metric (\ref{seed}) with the deformed parameters (\ref{new}) can be written  as $\bar{\mathcal{G}}_{ik}=8\pi ( T_{ik}+\mathcal{T}_{ik}) $. This leads to the following equations:
\begin{equation}
    8\pi (T_{0}^{0}+\mathcal{T}_{0}^{0})=-\frac{1}{r^{2}}+e^{-\beta}(\frac{1}{r^{2}}-\frac{\beta'}{r}),
\end{equation}
\begin{equation}
    8\pi (T_{1}^{1}+\mathcal{T}_{1}^{1})=-\frac{1}{r^{2}}+e^{-\beta}(\frac{1}{r^{2}}+\frac{\nu' +\alpha \zeta'}{r}),
\end{equation}
\begin{equation}
    8\pi (T_{2}^{2}+\mathcal{T}_{2}^{2})=\frac{1}{4}e^{-\beta}(2(\nu''+\alpha \zeta'')+(\nu'+\alpha \zeta')^{2}-\beta'(\nu'+\alpha \zeta')+2\frac{\nu' +\alpha \zeta' -\beta'}{r}),
\end{equation}
where $e^{\beta}=e^{\lambda}+\alpha \eta$ and  the prime sign denotes the partial derivative with respect to the radial coordinate $r$.

Considering the vacuum solution, specifically $T_{ik}=0$, which is known as the Schwarzschild solution, the Einstein field equations can be solved to reveal the emergence of the hairy Schwarzschild solution. This solution corresponds to the geometrical deformations described by the field equations. The line element for this solution is,

\begin{equation}\label{hairysch}
 	ds^{2}=-(1-\frac{2m}{r}+\alpha e^{-\frac{r}{m-\frac{\alpha l}{2}}})dt^{2}+(1-\frac{2m}{r}+\alpha e^{-\frac{r}{m-\frac{\alpha l}{2}}})^{-1}dr^{2}+r^{2}d\Omega^{2}.
 \end{equation}
Here $\alpha$ is coupling constant, $l$ is a new parameter with length dimension and associated with a primary hair of the black hole and $m$ is the mass of the black hole with relation to Schwarzschild mass $\mathcal{M}$,
\begin{equation}
	m=\mathcal{M}+\frac{\alpha l}{2}.
\end{equation}
The investigation of the influence of $\alpha$  and $l$ on various aspects such as geodesic motion, gravitational lensing, energy extraction, and thermodynamics has been extensively explored in the literature\cite{52y,53y,54y,55y,56y}. In this context, the present study aims to examine their impact on the change of the Bondi mass in the presence of gravitational waves within the hairy Kiselev background. The examination of the changes in the Bondi mass with respect to the coupling constant $\alpha$ and the length parameter $l$ is motivated by the dynamic and time-dependent nature of the Bondi mass, in contrast to the static nature of the ADM mass. The Bondi mass serves as a measure of mass on an asymptotically null slice or the energy densities on celestial spheres. 

In order to incorporate the gravitational memory effect at null infinity within the framework of the Bondi-Sachs formalism, it is deemed suitable to employ a transition to Eddington-Finkelstein coordinates.The outgoing Eddington-Finkelstein coordinates can be obtained by substituting the coordinate $t$ with $u= t-r^{*}$, thus defining this coordinate. 
This results in the transformation of the line element $(\ref{hairysch})$ which is given by
\begin{equation}\label{hairyschEF}
 	ds^{2}=-(1-\frac{2m}{r}+\alpha e^{-\frac{r}{m-\frac{\alpha l}{2}}})du^{2}-2dudr+r^{2}d\Omega^{2}.
\end{equation}

The exponential term  $\alpha e^{-\frac{r}{m-\frac{\alpha l}{2}}}$ diminishes as $r$ becomes large, making it impossible to detect its presence in the gravitational memory through its asymptotic behavior. However, the influence of the coupling constant  $\alpha$ and the length parameter $l$ can still be observed in the gravitational memory effects, as will be discussed shortly. To explore the gravitational memory effects in a broader context, we extended the solution of the hairy Schwarzschild black hole $(\ref{hairyschEF})$ to the surrounding hairy Kiselev solution \cite{yagoub}.

To obtain this solution one can consider the general from of the following line element
\begin{equation}\label{metricz}
    ds^{2}=-f(r)du^{2}-2dudr+r^{2}d\Omega^{2}
\end{equation}
The Einstein tensor by the metric $(\ref{metricz})$ can be defined as
\begin{equation}\label{e1}
 G^{0}_{0}=G^{1}_{1}=\frac{1}{r^{2}}(f'r-1+f),   
\end{equation}
\begin{equation}\label{e2}
 G^{2}_{2}=G^{3}_{3}=\frac{1}{r^{2}}(f'r+\frac{1}{2}r^{2}f''),  
\end{equation}
where the prime sign indicates the derivative with respect to $r$ coordinate. The total energy-momentum tensor is expressed as a sum of $\mathcal{T}_{ik}$ related to minimal geometrical deformations and $T_{ik}$ related to the surrounding fluid, given by $\Bar{T}_{ik}=T_{ik}+\alpha \mathcal{T}_{ik}$. In this case, the conditions $T_{;k}^{ik}=\mathcal{T}_{;k}^{ik}=0$ are not imposed, and only the Bianchi identity $\Bar{T}_{;k}^{ik}=0$ is required. By examining equations $(\ref{e1})$, $(\ref{e2})$, and metric $(\ref{metricz})$, we observe that the Einstein tensor exhibits the same symmetry for $\Bar{T}_{;k}^{ik}=0$, specifically $\Bar{T}_{0}^{0}=\Bar{T}_{1}^{1}$ and $\Bar{T}_{2}^{2}=\Bar{T}_{3}^{3}$. 

One can define the appropriate energy-momentum tensor for surrounding fluid as
\begin{equation}\label{e3}
    T^{0}_{0}=-\rho(r) ~~~~~~~and ~~~~~~~
   T^{i}_{k}=-\rho(r)[-\eta(1+3\zeta)\frac{r^{i}r_{k}}{r^{n}r_{n}}+\zeta \delta^{i}_{k}] .
\end{equation}

 The energy-momentum tensor $(\ref{e3})$ reveals that the spatial distribution is directly related to the temporal component, representing the energy density $\rho$ with unspecified parameters $\eta$ and $\zeta$ that are influenced by the internal configuration of the surrounding fields. The isotropic averaging across all angles leads to
\begin{equation}
   \langle T^{i}_{k} \rangle =\frac{\eta}{3} \rho \delta^{i}_{k}=P \delta^{i}_{k}
\end{equation}
where we used $\langle r^{i} r_{k} \rangle =\frac{1}{3}\delta ^{i}_{k}r_{n}r^{n}$. Therefore, the barotropic equation of state for the surrounding fluid can be given by
\begin{equation}\label{rho}
    P(r)=\omega \rho(r),  ~~~~~~~~\omega=\frac{\eta}{3}.
\end{equation} 
The pressure and constant equation of state parameter of the surrounding field are denoted as $P(r)$ and $\omega$, respectively. It is important to note that the source $T_{ik}$ associated with the surrounding field must possess the same symmetries as $\Bar{T}_{ik}$. This is because the source $\mathcal{T}_{ik}$, which represents the geometrical deformation source, has the same symmetries as $\mathcal{T}_{0}^{0}=\mathcal{T}_{1}^{1}$ and $\mathcal{T}_{2}^{2}=\mathcal{T}_{3}^{3}$. In simpler terms, we can express this as $T_{0}^{0}=T_{1}^{1}$ and $T_{2}^{2}=T_{3}^{3}$. Consequently, the free parameter $\zeta$ of the energy-momentum tensor $T_{ik}$ for the surrounding fluid can be determined as follows
\begin{equation}\label{zeta}
    \zeta=-\frac{1+3\omega}{6\omega}
\end{equation}
Upon substituting equations $(\ref{zeta})$ and $(\ref{rho})$ into equation $(\ref{e3})$, we obtain the surrounding energy momentum tensor with following non vanishing components
\begin{equation}
    T_{0}^{0}=T_{1}^{1}=-\rho ~~~~~~~~~~ and ~~~~~~~~ T_{2}^{2}=T_{3}^{3}=\frac{1}{2}(1+3\omega)\rho .
\end{equation}
By inserting total energy momentum tensor $\Bar{T}_{ik}$ into Einstein field equation we have
\begin{equation}
    f(r)= g(r)-\frac{\alpha l}{r}+\alpha e^{-\frac{r}{m-\frac{\alpha l}{2}}}
\end{equation}
Where $g(r)$ is given by
\begin{equation}
    g(r)=1-\frac{2\mathcal{M}}{r}+\frac{N}{r^{3\omega+1}}.
\end{equation}

Therefore the line element for this extended solution is expressed as follows. 

\begin{equation}\label{metrichk}
 		ds^{2}=-(1-\frac{2m}{r}+\frac{N}{r^{3\omega+1}}+\alpha e^{-\frac{r}{m-\frac{\alpha l}{2}}})du^{2}-2dudr+r^{2}d\Omega^{2},
 \end{equation}
where $N$ is the surrounding field structure parameter and $\omega$ is the constant equation of the state parameter of the surrounding field. The weak energy condition, a significant concept in our investigation of this model, is expressed as follows: \cite{yagoub} 
\begin{equation}\label{32}
    \rho(r)=\frac{3\omega N}{r^{3(\omega+1)}}\ge 0.
\end{equation}
 Here, $\rho$ represents the energy density of the surrounding field.   As previously mentioned, the exponential term in the metric  $(\ref{metrichk})$ becomes negligible for larger values of $r$. However, the coupling constant $\alpha$ and length parameter $l$ remain significant within the black hole mass  $m=\mathcal{M}+\frac{\alpha l}{2}$, which decreases with an order of  $\mathcal{O}(r^{-1})$ in the spacetime metric, specifically as $\frac{2m}{r}$.
 
 Furthermore, we are particularly interested in a dust-like field with the  constant equation of the state parameter $\omega=0$ in the hairy-surrounded Kiselev solution. This is because the parameter $N$, which represents the surrounding field structure in the metric as $\frac{N}{r}$, can be considered to have an order of  $\mathcal{O}(r^{-1})$. This is relevant for studying the changes in Bondi mass at null infinity within the Bondi-Sachs formalism. Consequently, in this section, we do not consider the parameter $\omega \not= 0$, and solely focus on the parameter $\omega = 0$. With these considerations, the line element that is of interest to us is as follows 
\begin{equation}\label{metricRRR}
 ds^{2}=-(1-\frac{2m}{r}+\frac{N}{r} + \alpha e^{-\frac{r}{m-\frac{\alpha l}{2}}})du^{2}-2dudr+r^{2}d\Omega^{2}.
 \end{equation}
\section{Bondi-Sachs formalism}
This section is dedicated to establishing our notation and examining the Bondi formalism for asymptotically Minkowskian metrics. Initially, we focus on the off-shell metric in Bondi gauge. In this context, we work with the Lorentzian space-time manifold characterized by the Bondi retarded coordinates $(u, r, x^{A})$. Here, $u$ represents the retarded time, $r$ denotes the radius, and $x^{A}~(A=1,2)$ refer to the coordinates on a celestial sphere at future null infinity. The manifold is defined by a metric $ds^{2}=g_{\mu\nu}dx^{\mu}dx^{\nu}$ where the components are chosen to adhere to the Bondi gauge.
\begin{equation}
    g_{rr}=g_{Ar}=0, ~~~~~~~~~~~~~~~~~ \partial_{r}\det(r^{-2}g_{AB})=0.
\end{equation}
Therefore one can write such a metric as follows
 \begin{equation}\label{bondi-sachs}
 	ds^{2}=-e^{2\beta}Fdu^{2}-2e^{2\beta}dudr+r^{2}h_{AB}(dx^{A}-\frac{U^{A}}{r^{2}}du)(dx^{B}-\frac{U^{B}}{r^{2}}du).
 \end{equation}
  where $h_{AB}$ is given by 
 \begin{equation}\label{hAB}
 	h_{AB}=\gamma_{AB}+C_{AB}r^{-1}+D_{AB}r^{-2}+\mathcal{O}(r^{-3}).
 \end{equation} 
In this scenario, let $A,B,...=1,2$ and $x^{1}=\theta$ , $x^{2}=\phi$. The metric on the unit 2 sphere, denoted as $\gamma_{AB}$, is selected as $\gamma_{AB}dx^{A}dx^{B}=d\theta^{2}+\sin{\theta}^{2}d\phi^{2}$.
We denote the shear tensor by $C_{AB}=C_{AB}(u,x^{A})$ and the leading order tensor induced by the GW pulse by $D_{AB}=D_{AB}(u,x^{A})$ which are  are raised with the inverse of $\gamma_{AB}$. 
The metric $h_{AB}$ will ultimately account for the impact of gravitational radiation, a vector field $U^{A}$ will ultimately convey details about angular momentum, and two functions $\beta$ and $F$ where the latter will ultimately detect the mass of the source. Prior to being determined, all these quantities are arbitrary functions on space-time. In order to guarantee that the metric $(\ref{bondi-sachs})$ truly exhibits asymptotically flat behavior, an additional boundary condition is enforced. For large $r$ we have $h_{AB}(u,r,x^{A}) \equiv \gamma_{AB}(x^{A})$ where $\gamma_{AB}(x^{A})$ is static and radius-independent metric of a unit sphere.

The Bondi framework implies that Einstein’s equations can be solved by a perturbative approach in $1/r$, that is, by an expansion around null infinity. Therefore, the arbitrary functions of $(\ref{bondi-sachs})$ are restricted by vacuum dynamics to asymptotic expansions. Utilizing the metric $(\ref{bondi-sachs})$, one can employ  it to solve the gravitational field equations to derive the subsequent outcomes from the effective Einstein field equations. The equation for the hypersurface $E_{\mu\nu}=R_{\mu\nu}-\frac{1}{2}g_{\mu\nu}R-8\pi G_{0}\bar{T}_{\mu\nu}=0$ provides us with the equation for $E^{u}_{r}=0$, which ultimately results in  
\begin{equation}
\partial_{r}\beta=\frac{r}{16}h^{AC}h^{BD}\partial_{r}h_{AB}\partial_{r}h_{CD}+2\pi \bar{T}{rr}.
\end{equation}

In the Bondi-Sachs null coordinate system, the source $\bar{T}_{rr}$ becomes zero for the specific scenario under consideration. Consequently, one can determine the value of $\beta$ as follows 

\begin{equation}\label{c1}
\beta= \frac{\beta_{2}}{r^{2}}+\mathcal{O}(r^{-3}),
\end{equation}
where $\beta_{2}=-\frac{1}{32}C^{2}$ and $C^{2}=C_{AB}C^{AB}$. We use the hypersurfaces $E^{u}_{r}=0$ and $E_{u}^{u}=0$ to determine $U^{A}$ and $F$, respectively  \cite{771,772}.

Hence, by taking into account the aforementioned factors and expanding the function  $F$ to leading order one can obtain
 \begin{equation}\label{c2}
 	F=1-\frac{2M}{r}-\frac{2F_{2}}{r^{2}}+\mathcal{O}(r^{-3}).
 \end{equation}
The Bondi mass, denoted as $M$, can be determined through the balance equation \cite{772}
 \begin{equation}\label{m}
     \dot{M}=\frac{1}{4}D_{A}D_{B}N^{AB}-\frac{1}{8}N_{AB}N^{AB}-4\pi r^{2}\bar{T}_{uu},
 \end{equation}
where over dot is denoted for derivative with respect to $u$.
The detailed calculation of $(\ref{m})$ can be found in reference\cite{772}. Utilizing the condition$E^{u}_{r}=0$, it follows that  \cite{772}. 
 \begin{equation}\label{c3}
 	U^{A}=U^{0A}+\frac{\sigma(u,x^{A})}{r}+\mathcal{O}(r^{-2}),
 \end{equation} 
the term $ \sigma^{A} \equiv\sigma(u,x^{A})$ is given by
\begin{equation}
    \sigma^{A}=-\frac{2}{3}L^{A}+\frac{1}{16}D^{A}(C_{BC}C^{BC})+\frac{1}{2}C^{AB}D^{C}C_{BC},
\end{equation}
where $L^{A}(u,\theta)$ is the angular momentum aspect, and the derivative of it with respect to $u$ is defined by 
\begin{equation}
    \dot{L}_{A}=D_{A}M+\frac{1}{2}D^{B}D_{[A}D^{C}C_{B]C}+\frac{1}{4}D_{B}(N^{BC}C_{AC})+\frac{1}{2}D_{B}N^{BC}C_{AC}.
\end{equation}
We denote $D_{A}$ for the covariant derivative with respect to $\gamma_{AB}$ (spherical Levi-Civita connection) and $A$, $B$ are lowered and raised by the static metric $\gamma_{AB}$. Note that, the densities of energy and angular momentum on celestial spheres are measured by the Bondi mass $M$ and the angular momentum aspect $L_{A}(u,\theta)$.

Finally, by using equations $(\ref{c1})$, $(\ref{c2})$ and $(\ref{c3})$, we obtain the on-shell metric
 \begin{eqnarray}\label{16}
 	ds^{2}=-(1-\frac{2M}{r}-\frac{2F_{2}}{r^{2}})du^{2}-2(1-\frac{\beta_{2}}{r^{2}})dudr+\\\nonumber
 	(r^{2}\gamma_{AB}+rC_{AB}+D_{AB})dx^{A}dx^{B}\\\nonumber
 	-2(\gamma_{AB}U^{0A}+\frac{\gamma_{AB}\sigma^{A}+C_{AB}U^{0A}}{r}+\frac{C_{AB}\sigma^{A}+D_{AB}U^{0A}}{r^{2}})dudx^{B}+\\\nonumber
 	\gamma_{AB}\frac{U^{0A}U^{0B}}{r^{2}}du^{2}+\mathcal{O}(r^{-3}).
 \end{eqnarray}

 \section{Kiselev black hole and Bondi-Sachs formalism}
In This section, we incorporate the GW components, before analyzing it within the framework of the Bondi-Sachs formalism.
 Since we are interested in asymptotic behaviour of the metric $(\ref{metricRRR})$ as we will explain in the following, the line element $(\ref{metricRRR})$ for large $r$ can be written as
\begin{equation}\label{metricRR}
 ds^{2}=-(1-\frac{2m}{r}+\frac{N}{r} )du^{2}-2dudr+r^{2}d\Omega^{2}.
 \end{equation}
We adopt the Bondi-Sachs formalism at null infinity to study the gravitational memory effects in the presence of a gravitational wave (GW) pulse propagating on the background metric $(\ref{metricRR})$, which is a solution of the effective gravitational field equations $(\ref{effectiveGR})$. These equations are equivalent to the generalized GR equations, as shown in \cite{771}. We apply  the Bondi-Sachs approach accordingly to account for the GW pulse, which reaches null infinity and changes the Bondi mass. We note that the background metric $(\ref{metricRR})$ is static and does not exhibit any Bondi mass dynamics in the absence of the GW pulse.

We impose the GW components on the null coordinate metric $(\ref{metricRR})$, expressed in the Bondi-Sachs form to achieve our objective. Thus, we obtain
 \begin{eqnarray}\label{perturbationmetric}
 ds^{2}=-du^{2}-2dudr+(\frac{2m}{r}-\frac{N}{r}+\frac{2M_{B}(u,\theta)}{r}+\frac{E(u,\theta)}{r^{2}})du^{2}-\\\nonumber 2(
 \frac{b_{1}(u,\theta)}{r}+\frac{b_{2}(u,\theta)}{r^{2}})dudr
 -2r^{2}U(u,r,\theta)dud\theta+r^{2}h_{AB}dx^{A}dx^{B}....,
 \end{eqnarray}
  where $h_{AB}$ is given by $(\ref{hAB})$. We denote the shear tensor by $C_{AB}=C_{AB}(u,x^{A})$ and the leading order tensor induced by the GW pulse by $D_{AB}=D_{AB}(u,x^{A})$ which are  are raised with the inverse of $\gamma_{AB}$.
  This term is disregarded  in our analysis. The GW pulse affects the Bondi mass $M_{B}(u,\theta)$, which is dynamic. The metric $(\ref{perturbationmetric})$ also contains other dynamical terms, which are perturbations of the background metric caused by the GW pulse.
 
We use the Bondi-Sachs metric to solve the gravitational field equation in a perturbative manner. We have introduced the off-shell solution of the Bondi metric with the line element $(\ref{bondi-sachs})$. To solve the gravitational field equation in a perturbative approach, we insert the metric $(\ref{bondi-sachs})$ into equation $(\ref{effectiveGR})$ and consider the large $r$ limit at null infinity. We then compare the solved metric with the metric $(\ref{perturbationmetric})$ order by order to find the change in the Bondi mass $\ M_{B}(u,\theta)$. We will present this analysis shortly, but first, we will explain some aspects of the Bondi shear tensor and the News tensor.

The GW pulse in our scenario is characterized by the tensorial News $N_{AB}$, which is given by
 \begin{equation}\label{c}
    N_{AB} =\partial_{u}C_{AB}= A_{0} sech^{2}(u) Y 
     \begin{pmatrix}
     1&0\\
     0&-\sin^{2}\theta
     \end{pmatrix}.
 \end{equation}
The GW pulse in our scenario is characterized by the News tensor  $N_{AB}$, which is defined by the $sech^{2}u$ function. This particular choice  is compatible with the upcoming section and signifies the GW pulse's impact on the background geometry. In this context, the selection of $N_{AB}$ in the aforementioned format is based on its mathematical simplicity and its alignment with the definition provided in the subsequent section.  The News tensor  $N_{AB}$ represents the gravitational degrees of freedom. To ensure that the GW pulse serves as a minor disturbance to the hairy black hole solution $(\ref{metricRR})$, we adjust the value of  $A_{0}$. Subsequently, we utilize equation $(\ref{c})$ to calculate the shear tensor $C_{AB}$.
\begin{equation}
     C_{AB}=A_{0}Y
     \begin{pmatrix}
          1&0\\
     0&-\sin^{2}\theta
     \end{pmatrix}\int_{-\infty}^{+\infty}sech^{2}(u)du=-2Y\begin{pmatrix}
          A_{0}&0\\
     0&-A_{0}\sin^{2}\theta
     \end{pmatrix},
 \end{equation}
 where $Y$ is the spin-weighted harmonic which is defined by $Y=\frac{3}{4}\sqrt{\frac{5}{6\pi}}\sin^{2}\theta$.

Now, for our aim, by comparing $(\ref{16})$ with the metric $(\ref{perturbationmetric})$ we have
 \begin{equation}\label{compare1}
 	M_{B}(u,\theta)=M-m+\frac{N}{2}.
 \end{equation}
 \begin{equation}\label{compare2}
 E(u,\theta)=-2(\frac{1}{6}(D_{A}L^{A})+\frac{1}{8}(D_{A}C^{AB})(D_{D}C_{B}^{D}))-\frac{1}{16}C^{2}.
 \end{equation}
 \begin{equation}\label{compare3}
 b_{2}(u,\theta)=\frac{1}{32}C^{2} ~~~~~~~,~~~~~~C^{2}=C_{AB}C^{AB}.
 \end{equation}
 \begin{equation}\label{compare4}
 b_{1}(u,\theta)=0.
 \end{equation}
 \begin{eqnarray}\label{compare5}
 U(u,r,\theta)=\frac{\gamma_{AB}U^{0A}}{r^{2}}+\frac{\gamma_{AB}(-\frac{2}{3}L^{A}+\frac{1}{16}D^{A}C^{2}+\frac{1}{2}C^{AB}D^{C}C_{BC})+C_{AB}U^{0A}}{r^{3}}+\\\nonumber
 \frac{C_{AB}(-\frac{2}{3}L^{A}+\frac{1}{16}D^{A}C^{2}+\frac{1}{2}C^{AB}D^{C}C_{BC})+D_{AB}U^{0A}}{r^{4}}.
 \end{eqnarray}
 In the equation $(\ref{compare1})$, $M$ is the Bondi mass.   For the Kiselev solution and its asymptotic behavior,  $\bar{T}_{uu}=0$. By inserting News tensor from equation $(\ref{c})$ into $(\ref{m})$  one can obtain

\begin{equation}\label{mm}
   \dot{M}=\partial_{u}M= \frac{1}{4}A_{0}sech^{2}u(Y_{\theta \theta}+Y\cos{2\theta}-Y_{\theta}\cot{\theta}+2Y\cot^{2}\theta) -\frac{1}{4}A_{0}^{2}Y^{2}sech^{4}u.
\end{equation}
 We denote by $Y_{\theta}$ and $Y_{\theta \theta}$ the first and second derivatives of $Y$ with respect to $\theta$, respectively.
\begin{figure}[!tbp]
  \centering
  \begin{minipage}[b]{0.4\textwidth}
    \includegraphics[width=\textwidth]{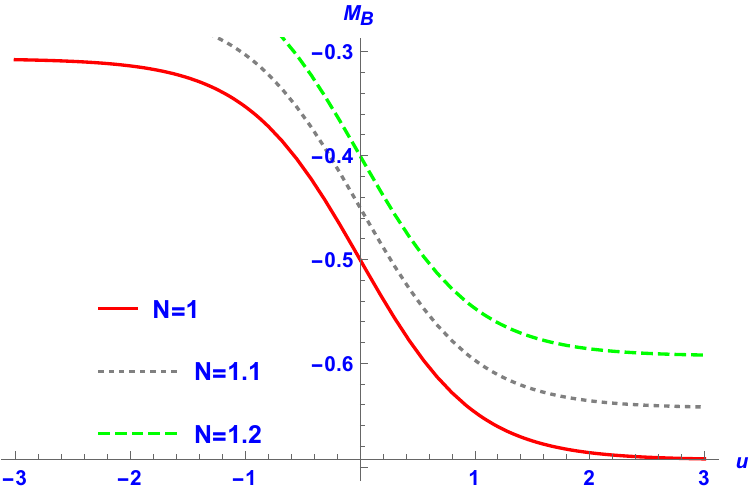}
    \caption{The variation of the change in the Bondi mass due to the GW pulse with respect to the null coordinate $u$ is shown for different choices of $N$.}
  \end{minipage}
  \hfill
  \begin{minipage}[b]{0.4\textwidth}
    \includegraphics[width=\textwidth]{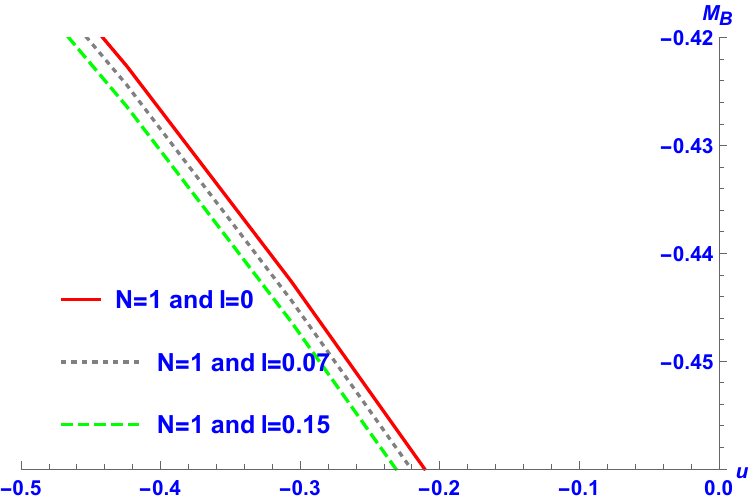}
    \caption{Variation of the change in the Bondi mass due to the GW pulse is presented with the null coordinate $u$, for different choices of the length parameter $l$}
  \end{minipage}
\end{figure}

 According to equation $(\ref{mm})$ $M$ can be obtained as
\begin{equation}
    M= \frac{1}{4}A_{0}(Y_{\theta \theta}+Y\cos{2\theta}-Y_{\theta}\cot{\theta}+2Y\cot^{2}\theta)\int du sech^{2}u-\frac{1}{4}A_{0}^{2}Y^{2}\int du sech^{4}u,
\end{equation}
which leads to 
\begin{equation}\label{M1}
    M= \frac{1}{4}A_{0}\tanh{u}(Y_{\theta \theta}+Y\cos{2\theta}-Y_{\theta}\cot{\theta}+2Y\cot^{2}\theta)-\frac{1}{4}A_{0}^{2}Y^{2}\tanh{u}(\frac{2}{3}+\frac{1}{3}sech^{2}u).
\end{equation}
Given the significance of the asymptotic behavior in analyzing the variation of the Bondi mass, it is possible to set the integral constant to zero in order to determine the value of $M$. Therefore, by using equations $(\ref{compare1})$ and $(\ref{M1})$, we can find the change in Bondi mass as
 \begin{equation}
    M_{B}(u,\theta)=-m+\frac{N}{2} + \frac{1}{4}A_{0}\tanh{u}(Y_{\theta \theta}+Y\cos{2\theta}-Y_{\theta}\cot{\theta}+2Y\cot^{2}\theta)-\frac{1}{4}A_{0}^{2}Y^{2}\tanh{u}(\frac{2}{3}+\frac{1}{3}sech^{2}u).
 \end{equation}
The evolution of the Bondi mass with variation in the surrounding field structure parameter $N$ has been illustrated in Figure $(1)$. Note that it is depicted for $M_{B}$ versus $u$ and for $\theta=\frac{\pi}{2}$ plane which leads to following equation for $M_{B}$
 \begin{equation}\label{bondialtered}
     M_{B}(u,\theta=\frac{\pi}{2})=-m+\frac{N}{2}-\frac{3}{8}\sqrt{\frac{5}{6 \pi}} A_{0}tanh(u)
 \end{equation}

Figure $(1)$ illustrates the variation of the Bondi mass and the coordinate $u$ for different values of $N$. It is evident that as $N$ increases, the increase in Bondi mass becomes more pronounced. According to equation  $(\ref{32})$, the weak energy condition requires a positive value for the surrounding field parameter N when the equation of state parameter $\omega$ is a positive constant. Conversely, in order for the solution$(\ref{metricRR})$ to have an event horizon, it is necessary to impose the condition $2m >N$. Hence, for the purposes of figure $(1)$, we have chosen  $m=1$ ,$A_{0}=1$ and $u_{0}=0$.

A hairy black hole is an intriguing black hole solution that goes beyond classical general relativity. It possesses new hair that originates from unknown fundamental fields. For example, for a rotating hairy black hole solution these unknown fundamental fields can be dark matter and dark energy \cite{54y}. These fields and any unknown source may  violate the No-Hair theorem of general relativity and suggest new theories of gravity and spacetime. In other words, the hairy black hole spacetime is a phenomenological model that can test the hypothesis of new fundamental fields. Therefore, we explore how the coupling constant $\alpha$ and the length parameter $l$ affect the Bondi mass in our model of interest. 

In doing so, we indicate $\alpha$ and $l$ in the equation $(\ref{bondialtered})$ explicitly by inserting $m=\mathcal{M}+\frac{\alpha l}{2}$ .
\begin{equation}
     M_{B}(u,\theta=\frac{\pi}{2})=-\mathcal{M}-\frac{\alpha l}{2}+\frac{N}{2}-\frac{3}{8}\sqrt{\frac{5}{6 \pi}} A_{0}tanh(u).
\end{equation}
The depiction of the alteration in Bondi mass for specific values of the length parameter $l$ is presented in figure $(2)$. In consideration of the energy condition $(\ref{32})$  and the requirement  $2m>N$, we assign $\mathcal{M}$ a value of 1 and $\alpha$ a value of  0.05. The pulse profile is determined by selecting $A_{0}=1$ and $u_{0}=0$.  Consequently, the trace of the hair of the Kiselev black hole at the null infinity can be determined by examining the change in Bondi mass. An identical assertion can be made regarding the coupling constant $\alpha$ and its impact on the Bondi mass.

\section{Gravitational memory effects and geodesics }
This section will present the analysis of the memory effect concerning the geodesic deviation between adjacent geodesics caused by the propagation of a gravitational wave. The geodesic separation serves as a measure of the extent of the displacement memory effect. Additionally, if the geodesics do not maintain a constant separation, a velocity memory effect can also be associated with these geodesics following the passage of the gravitational wave pulse.

In recent years, the examination of various effects related to the evolution of geodesics in exact plane GW spacetimes has been conducted  \cite{74,80,81,82,83,84,85}. To accomplish this, a Gaussian pulse has been utilized as the polarization (radiative) term in the line element of the plane GW spacetime. By numerically solving the geodesic equations, the displacement and velocity memory, which refer to the change in separation and velocity, respectively, can be determined as a result of the GW pulse passing through. Moreover, this formalism has been recently expanded to encompass alternative theories of gravity \cite{84,85}, with the aim of identifying unique characteristics of these theories within the displacement and velocity memory effects. 

In this study, we aim to investigate the dynamics of geodesics within the hairy Kiselev background, represented by the metric $(\ref{metrichk})$. We explore the resulting displacement and velocity memory effects by introducing a gravitational wave pulse into this background. The focus of our analysis is to understand the influence of the background geometry on these effects. 

To achieve this objective, first of all, we express the metric $(\ref{metrichk})$ at the subleading order in the following manner. 
\begin{equation}\label{metricsubleading}
 		ds^{2}=-(1-\frac{2m}{r}+\frac{N}{r^{3\omega+1}})du^{2}-2dudr+r^{2}d\Omega^{2}.
 \end{equation}

In order to achieve asymptotically flat behavior for the metric $(\ref{metricsubleading})$, we imposed a restriction on the parameter $\omega$, specifically $\omega > -\frac{1}{3}$. For our aim, we expressed the spacetime metric as the combination of the metric  $g_{\mu\nu}$ which is given by $(\ref{metricsubleading})$ and a perturbation metric $h_{\mu\nu}$. Therefore, we have
\begin{equation}
    ds^{2}=(g_{\mu\nu}+h_{\mu\nu})dx^{\mu}dx^{\nu}.
\end{equation}

\begin{figure}[!tbp]
  \centering
  \begin{minipage}[b]{0.4\textwidth}
    \includegraphics[width=\textwidth]{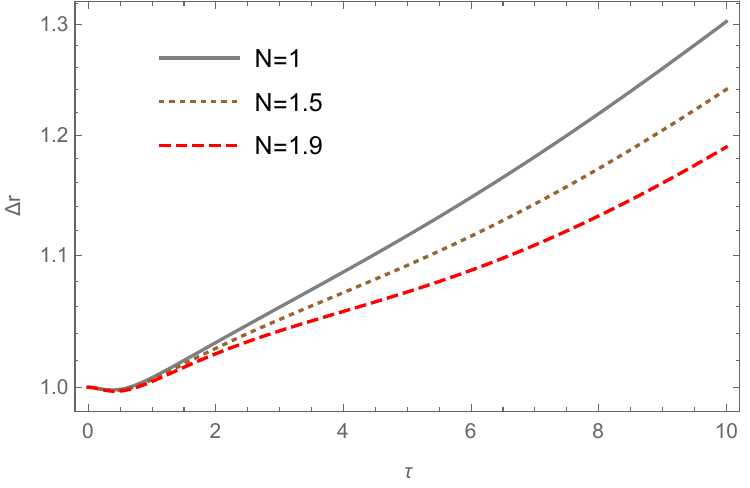}
    \caption{ Variation of the difference between the radial coordinates of the timelike geodesics, namely $\Delta r$, is presented against the proper time $\tau$ for different values of the Kiselev black hole with surrounding field parameter $N$ and $\omega=0$.}
  \end{minipage}
  \hfill
  \begin{minipage}[b]{0.4\textwidth}
    \includegraphics[width=\textwidth]{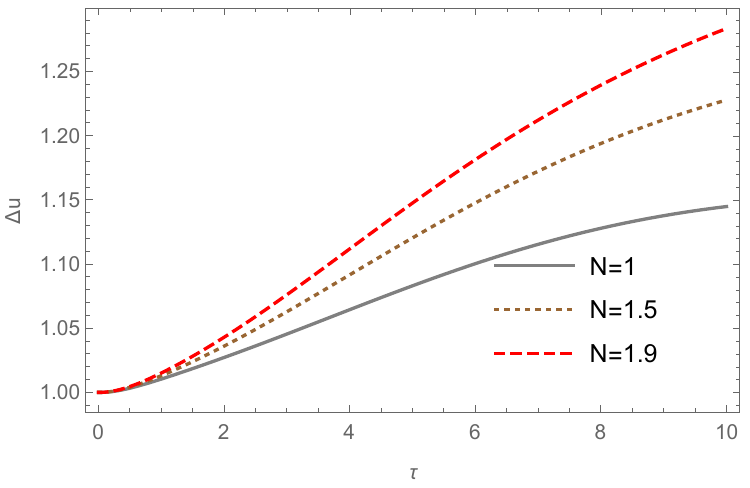}
    \caption{ Variation of the difference between the radial coordinates of the timelike geodesics, namely $\Delta u$, is presented against the proper time $\tau$ for different values of the Kiselev black hole with surrounding field parameter $N$ and $\omega=0$.}
  \end{minipage}
  \begin{minipage}[b]{0.4\textwidth}
    \includegraphics[width=\textwidth]{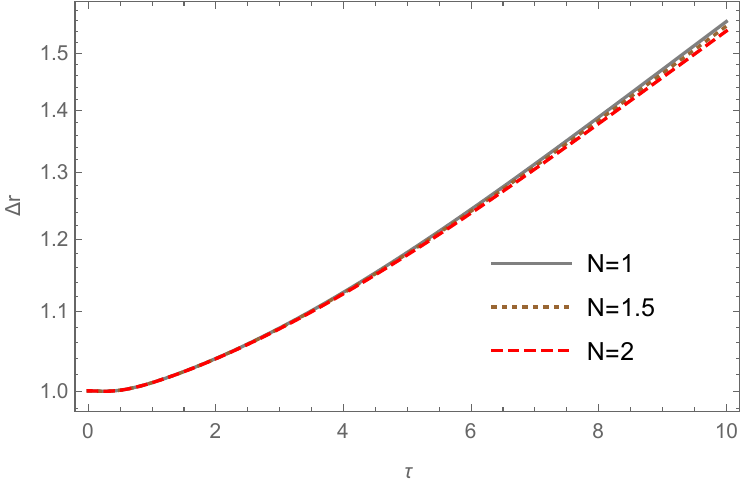}
    \caption{ Variation of the difference between the radial
coordinates of the timelike geodesics, namely $\Delta r$ has
been presented against the proper time $\tau$ for different
values of the Kiselev black hole with surrounding field parameter $N$ and $\omega=\frac{1}{3}$.}
  \end{minipage}
  \hfill
  \begin{minipage}[b]{0.4\textwidth}
    \includegraphics[width=\textwidth]{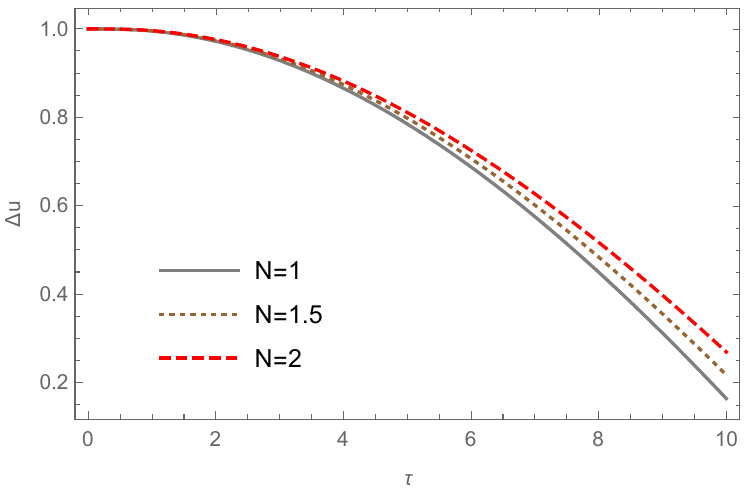}
    \caption{ Variation of the difference between null coordinate of
the two timelike geodesics, denoted by $\Delta u$ with respect
to the proper time $\tau$, for different values of Kiselev black hole with surrounding field parameter N and $\omega=\frac{1}{3}$.}
  \end{minipage}
  \begin{minipage}[b]{0.4\textwidth}
    \includegraphics[width=\textwidth]{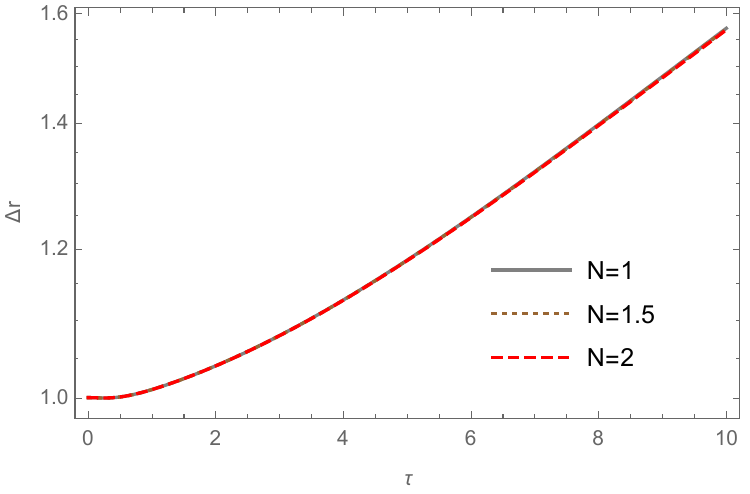}
    \caption{ Variation of the difference between the radial
coordinates of the timelike geodesics, namely $\Delta r$ has
been presented against the proper time $\tau$ for different
values of the Kiselev black hole with surrounding field parameter $N$ and $\omega=\frac{2}{3}$.}
  \end{minipage}
  \hfill
  \begin{minipage}[b]{0.4\textwidth}
    \includegraphics[width=\textwidth]{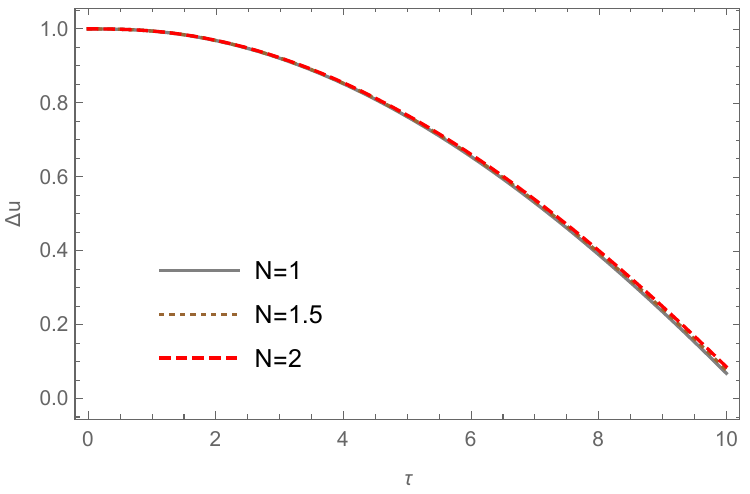}
    \caption{ Variation of the difference between null coordinate of
the two timelike geodesics, denoted by $\Delta u$ with respect
to the proper time $\tau$, for different values of Kiselev black hole with surrounding field parameter N and $\omega=\frac{2}{3}$.}
  \end{minipage}
\end{figure}

\begin{figure}[!tbp]
  \centering
  \begin{minipage}[b]{0.4\textwidth}
    \includegraphics[width=\textwidth]{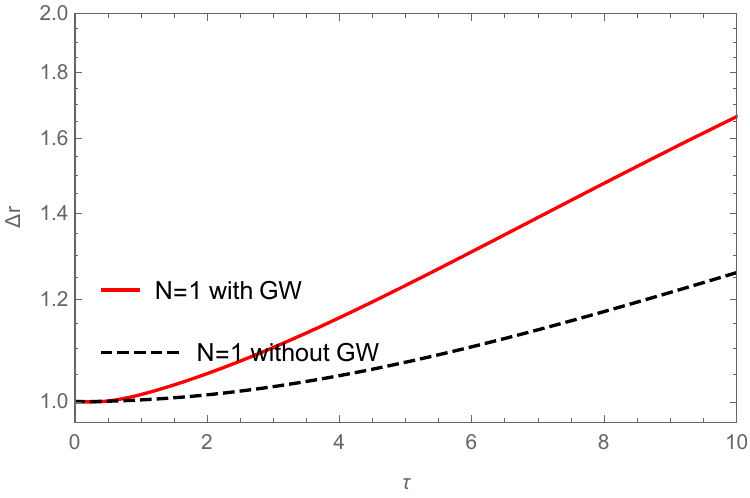}
    \caption{ The  Geodesic deviation $\Delta r$ without GW pulse and in the presence of GW pulse for the case $\omega=0$ with surrounding field parameter $N=1$.}
  \end{minipage}
  \hfill
  \begin{minipage}[b]{0.4\textwidth}
    \includegraphics[width=\textwidth]{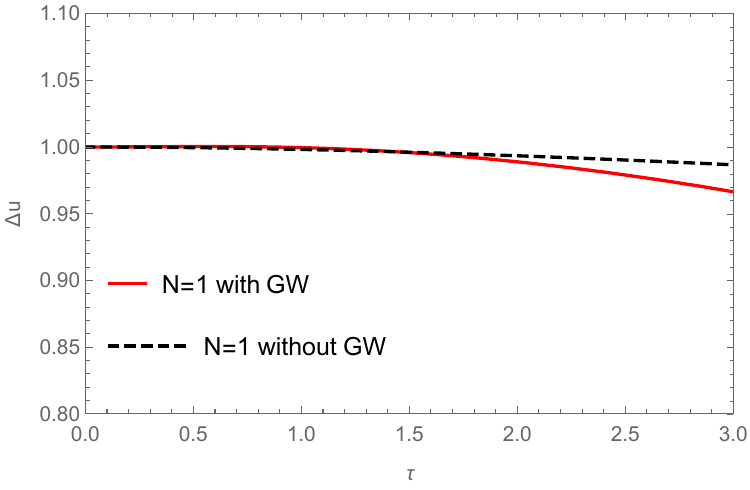}
    \caption{ The  Geodesic deviation $\Delta u$ without GW pulse and in the presence of GW pulse for the case $\omega=0$ with surrounding field parameter $N=1$.}
  \end{minipage}
  \begin{minipage}[b]{0.4\textwidth}
    \includegraphics[width=\textwidth]{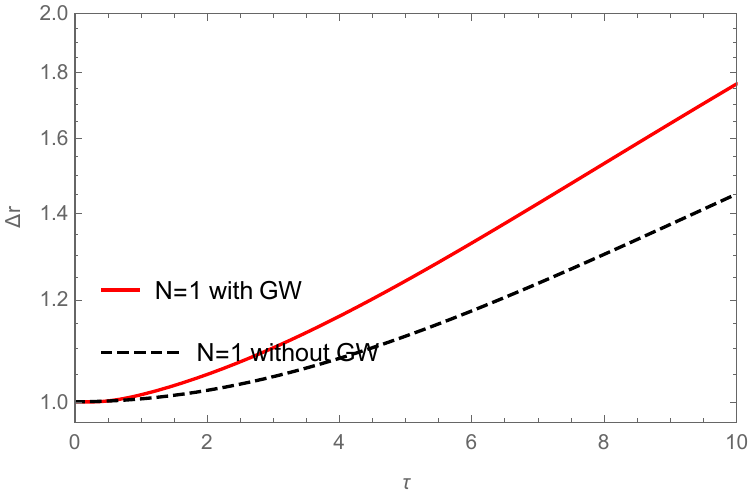}
    \caption{ The  Geodesic deviation $\Delta r$ without GW pulse and in the presence of GW pulse for the case $\omega=1/3$ with surrounding field parameter $N=1$.}
  \end{minipage}
  \hfill
  \begin{minipage}[b]{0.4\textwidth}
    \includegraphics[width=\textwidth]{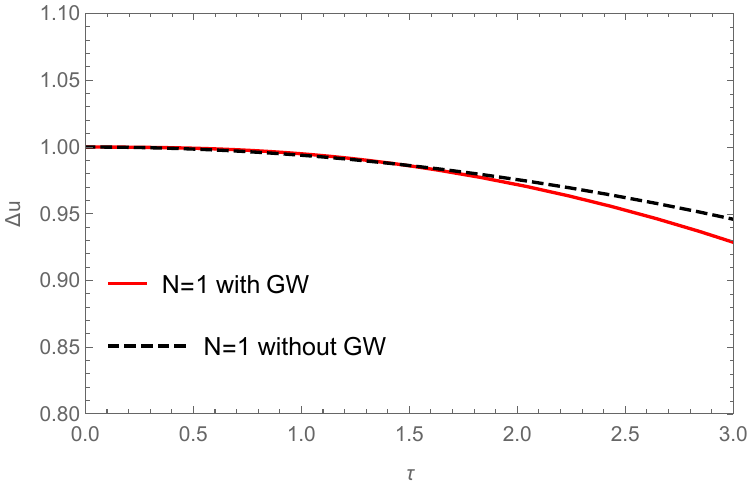}
    \caption{ The  Geodesic deviation $\Delta u$ without GW pulse and in the presence of GW pulse for the case $\omega=1/3$ with surrounding field parameter $N=1$.}
  \end{minipage}
  \begin{minipage}[b]{0.4\textwidth}
    \includegraphics[width=\textwidth]{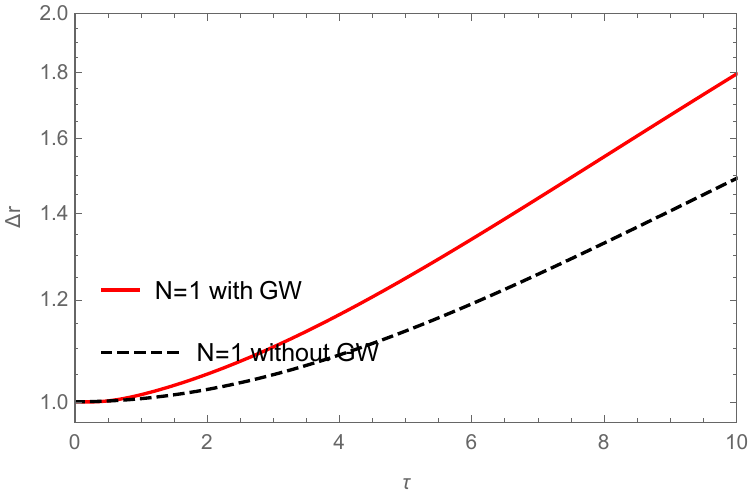}
    \caption{ The  Geodesic deviation $\Delta r$ without GW pulse and in the presence of GW pulse for the case $\omega=2/3$ with surrounding field parameter $N=1$.}
  \end{minipage}
  \hfill
  \begin{minipage}[b]{0.4\textwidth}
    \includegraphics[width=\textwidth]{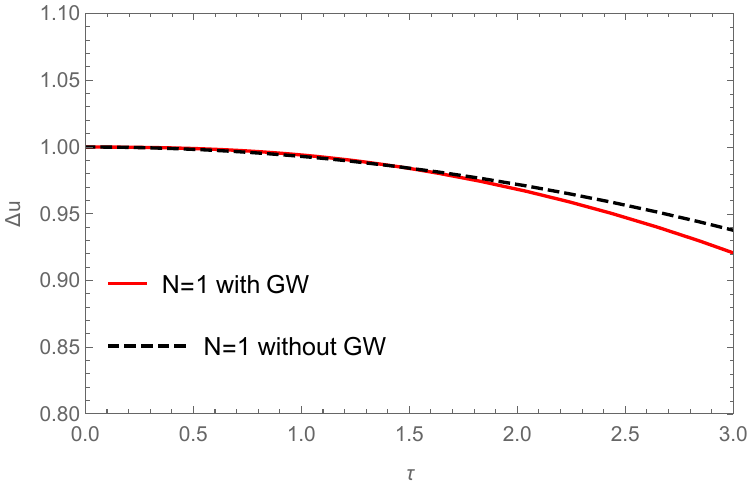}
    \caption{ The  Geodesic deviation $\Delta u$ without GW pulse and in the presence of GW pulse for the case $\omega=2/3$ with surrounding field parameter $N=1$.}
  \end{minipage}
\end{figure}
\begin{figure}[!tbp]
  \centering
  \begin{minipage}[b]{0.4\textwidth}
    \includegraphics[width=\textwidth]{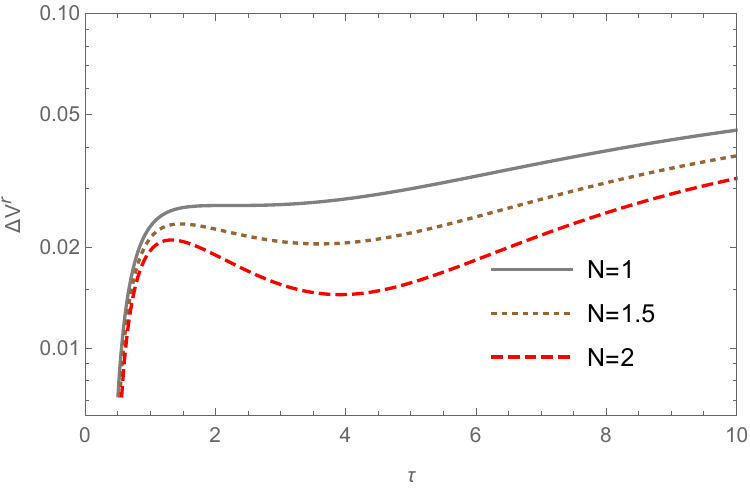}
    \caption{ Variation of  $\Delta V^{r}$ has
been presented against the proper time $\tau$ for different
values of the Kiselev black hole with surrounding field parameter $N$ and $\omega=0$.}
  \end{minipage}
  \hfill
  \begin{minipage}[b]{0.4\textwidth}
    \includegraphics[width=\textwidth]{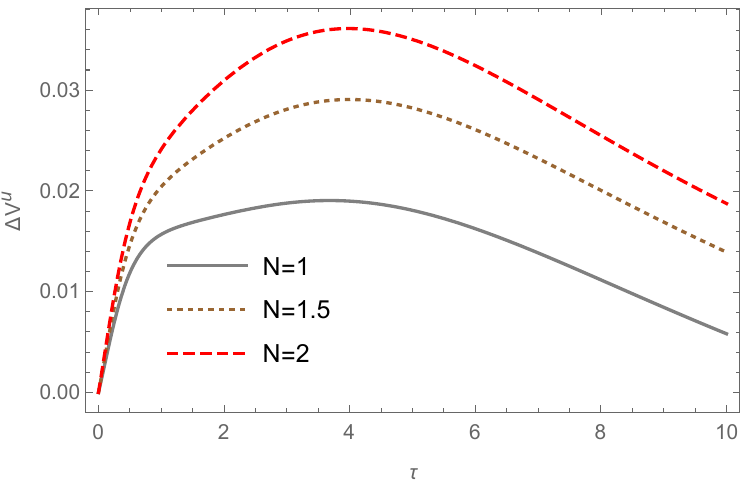}
    \caption{ Variation of  $\Delta V^{u}$ has
been presented against the proper time $\tau$ for different
values of the Kiselev black hole with surrounding field parameter $N$ and $\omega=0$.}
  \end{minipage}
  \begin{minipage}[b]{0.4\textwidth}
    \includegraphics[width=\textwidth]{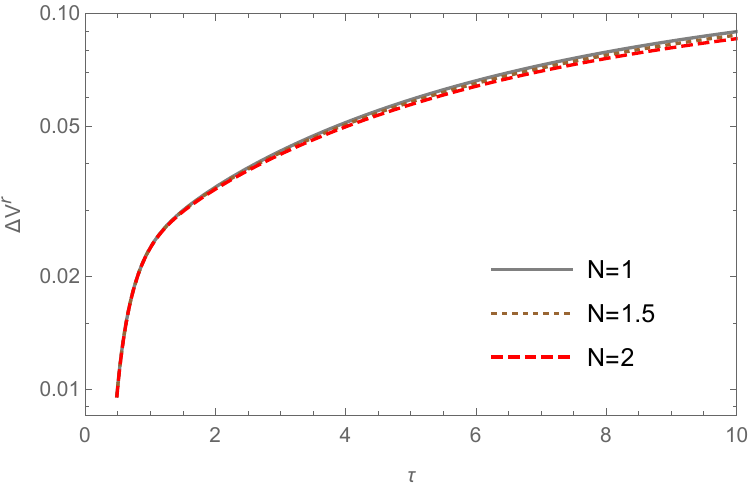}
    \caption{ Variation of  $\Delta V^{r}$ has
been presented against the proper time $\tau$ for different
values of the Kiselev black hole with surrounding field parameter $N$ and $\omega=1/3$.}
  \end{minipage}
  \hfill
  \begin{minipage}[b]{0.4\textwidth}
    \includegraphics[width=\textwidth]{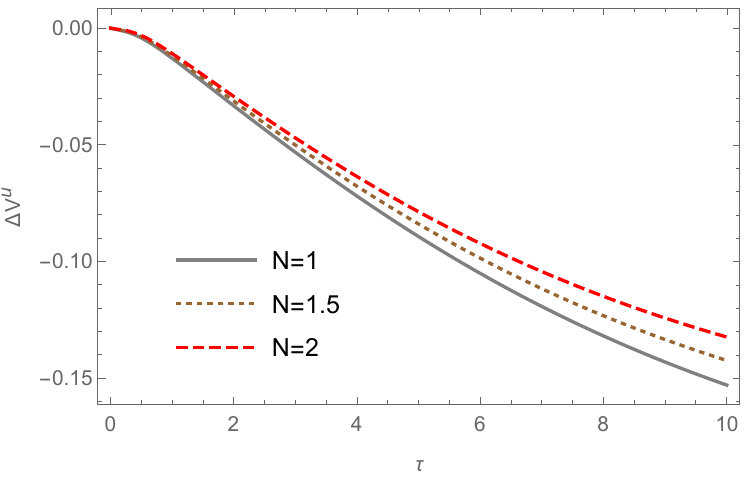}
    \caption{  Variation of  $\Delta V^{u}$ has
been presented against the proper time $\tau$ for different
values of the Kiselev black hole with surrounding field parameter $N$ and $\omega=1/3$.}
  \end{minipage}
  \begin{minipage}[b]{0.4\textwidth}
    \includegraphics[width=\textwidth]{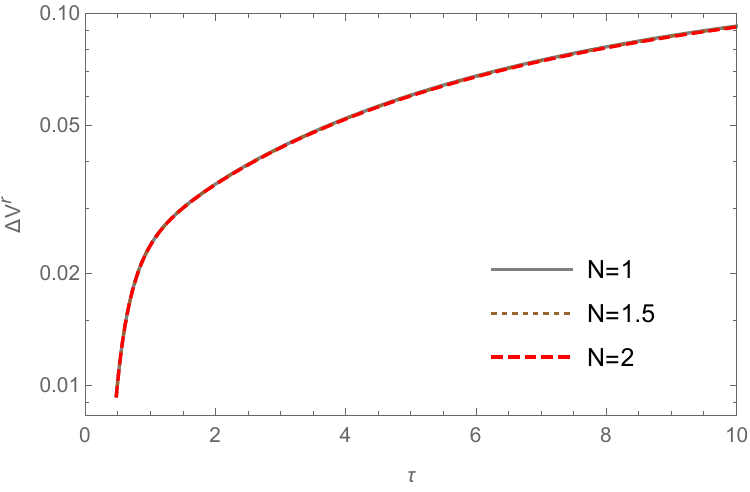}
    \caption{ Variation of  $\Delta V^{r}$ has
been presented against the proper time $\tau$ for different
values of the Kiselev black hole with surrounding field parameter $N$ and $\omega=2/3$.}
  \end{minipage}
  \hfill
  \begin{minipage}[b]{0.4\textwidth}
    \includegraphics[width=\textwidth]{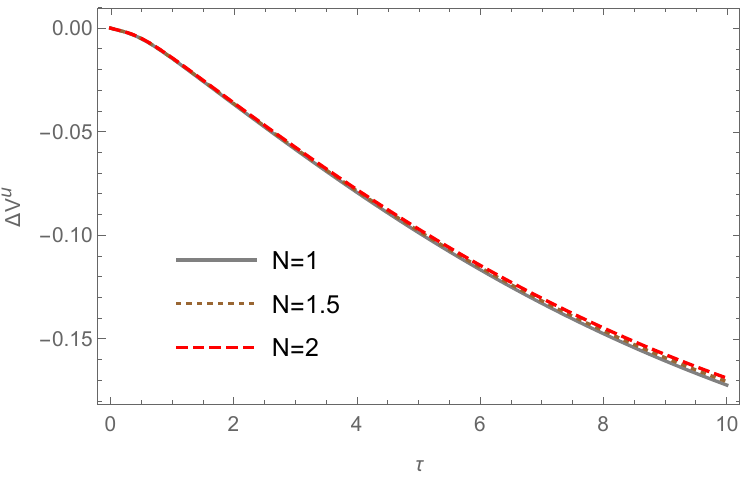}
    \caption{ Variation of  $\Delta V^{u}$ has
been presented against the proper time $\tau$ for different
values of the Kiselev black hole with surrounding field parameter $N$ and $\omega=2/3$.}
  \end{minipage}
\end{figure}

\begin{figure}[!tbp]
  \centering
  \begin{minipage}[b]{0.4\textwidth}
    \includegraphics[width=\textwidth]{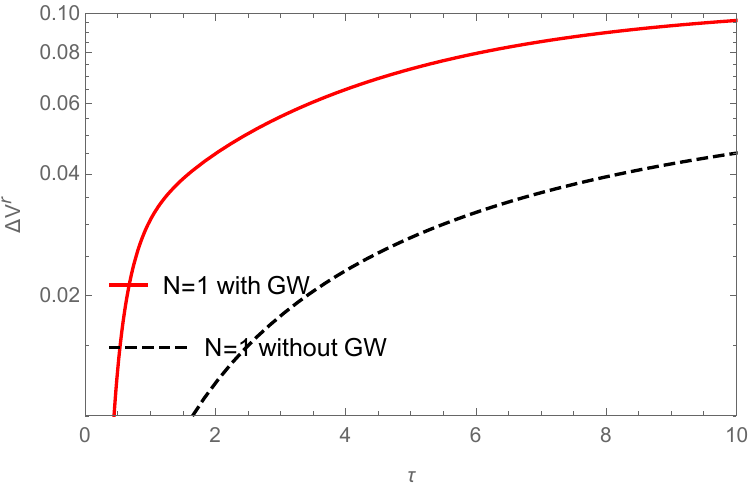}
    \caption{The variation of $\Delta V^{r}$ without GW pulse and in the presence of GW pulse for the case $\omega=0$ with surrounding field parameter $N=1$. }
  \end{minipage}
  \hfill
  \begin{minipage}[b]{0.4\textwidth}
    \includegraphics[width=\textwidth]{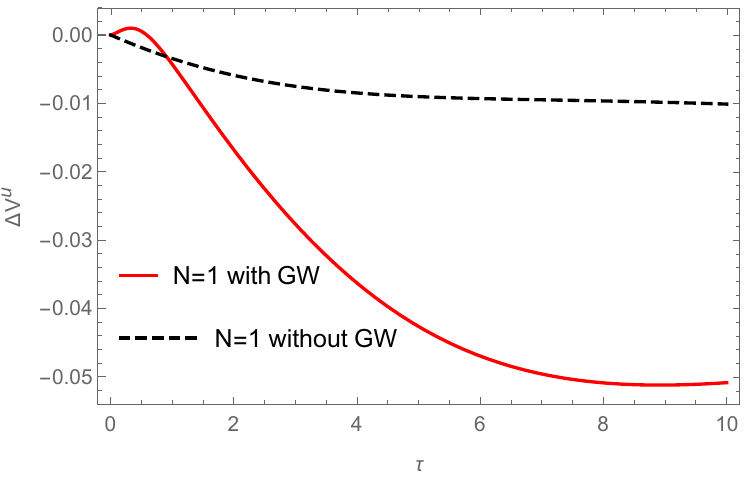}
    \caption{ The variation of $\Delta V^{u}$ without GW pulse and in the presence of GW pulse for the case $\omega=0$ with surrounding field parameter $N=1$.}
  \end{minipage}
  \begin{minipage}[b]{0.4\textwidth}
    \includegraphics[width=\textwidth]{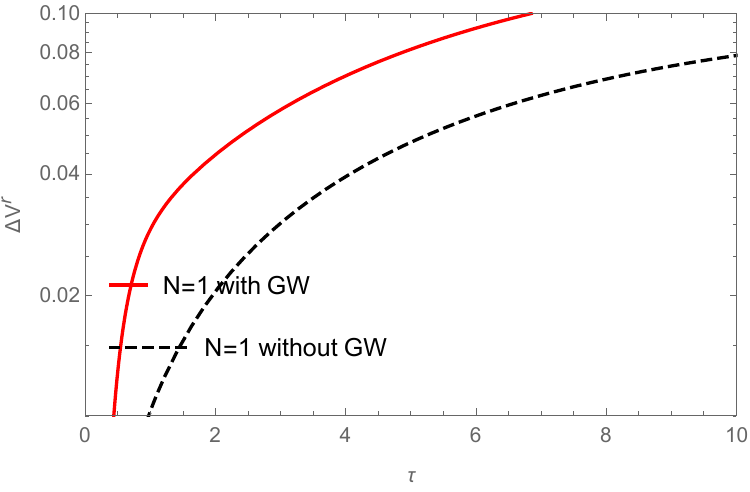}
    \caption{ The variation of $\Delta V^{r}$ without GW pulse and in the presence of GW pulse for the case $\omega=1/3$ with surrounding field parameter $N=1$.}
  \end{minipage}
  \hfill
  \begin{minipage}[b]{0.4\textwidth}
    \includegraphics[width=\textwidth]{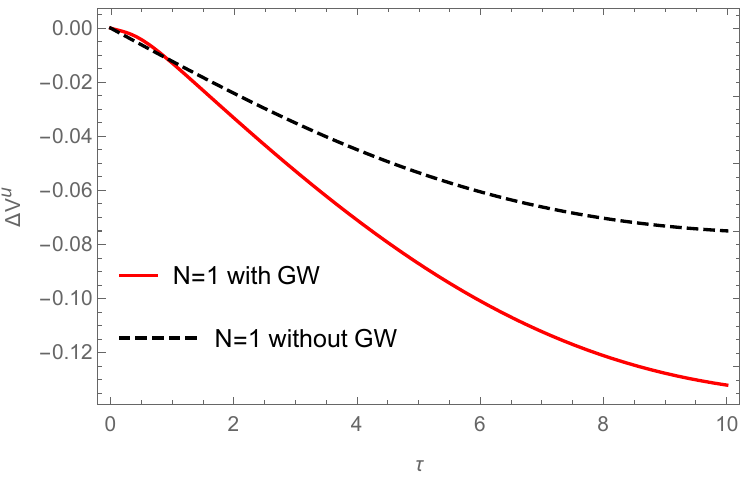}
    \caption{  The variation of $\Delta V^{u}$ without GW pulse and in the presence of GW pulse for the case $\omega=1/3$ with surrounding field parameter $N=1$.}
  \end{minipage}
  \begin{minipage}[b]{0.4\textwidth}
    \includegraphics[width=\textwidth]{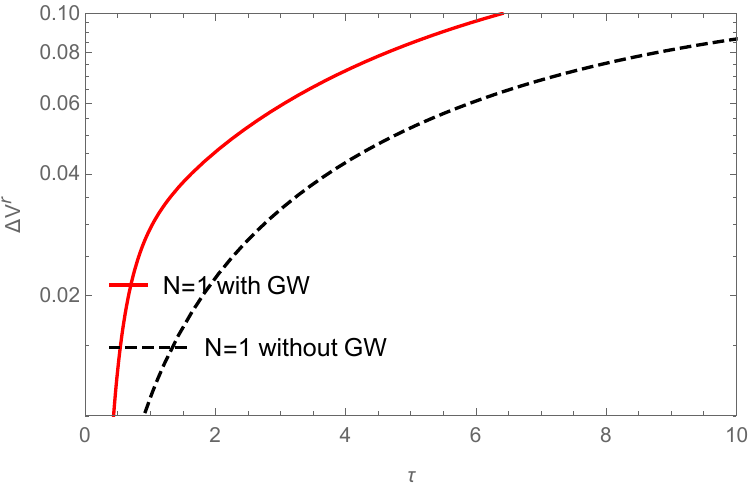}
    \caption{ The variation of $\Delta V^{r}$ without GW pulse and in the presence of GW pulse for the case $\omega=2/3$ with surrounding field parameter $N=1$.}
  \end{minipage}
  \hfill
  \begin{minipage}[b]{0.4\textwidth}
    \includegraphics[width=\textwidth]{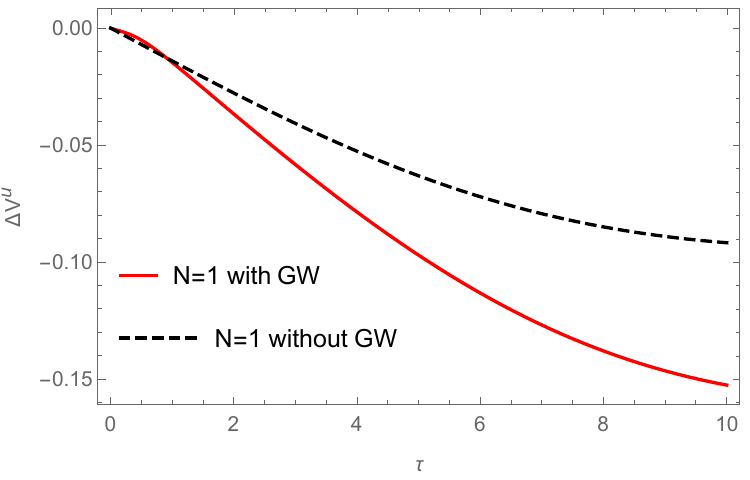}
    \caption{ The variation of $\Delta V^{u}$ without GW pulse and in the presence of GW pulse for the case $\omega=2/3$ with surrounding field parameter $N=1$.}
  \end{minipage}
\end{figure}

\begin{figure}[!tbp]
  \centering
  \begin{minipage}[b]{0.4\textwidth}
    \includegraphics[width=\textwidth]{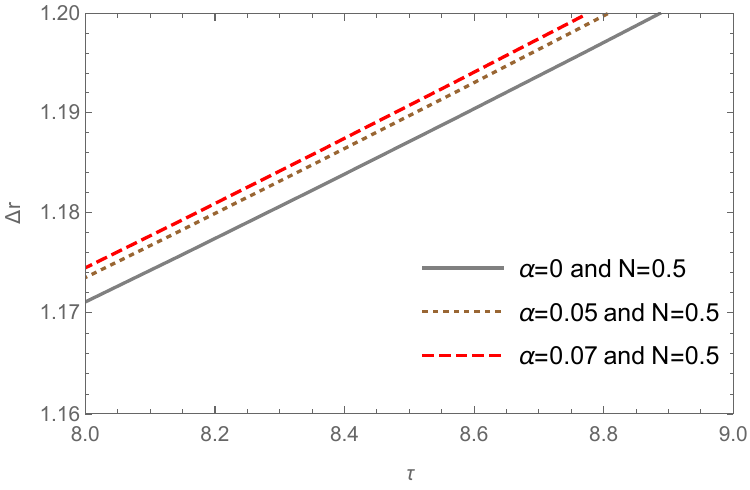}
    \caption{The variation of $\Delta r$ in the presence of a GW pulse for different values of $\alpha$ in the case $\omega=0$ with surrounding field parameter $N=0.5$. }
  \end{minipage}
  \hfill
  \begin{minipage}[b]{0.4\textwidth}
    \includegraphics[width=\textwidth]{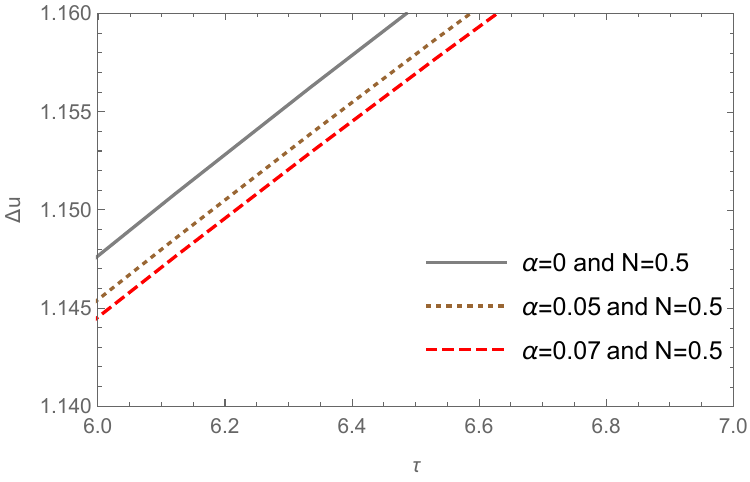}
    \caption{ The variation of $\Delta u$ in the presence of a GW pulse for different values of $\alpha$ in the case $\omega=0$ with surrounding field parameter $N=0.5$.}
  \end{minipage}
  \begin{minipage}[b]{0.4\textwidth}
    \includegraphics[width=\textwidth]{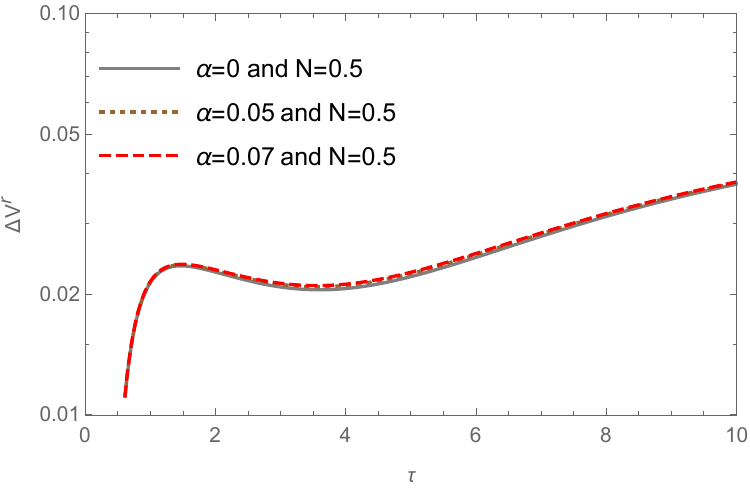}
    \caption{ The variation of $\Delta V^{r}$ in the presence of a GW pulse for different values of $\alpha$ in the case $\omega=0$ with surrounding field parameter $N=0.5$.}
  \end{minipage}
  \hfill
  \begin{minipage}[b]{0.4\textwidth}
    \includegraphics[width=\textwidth]{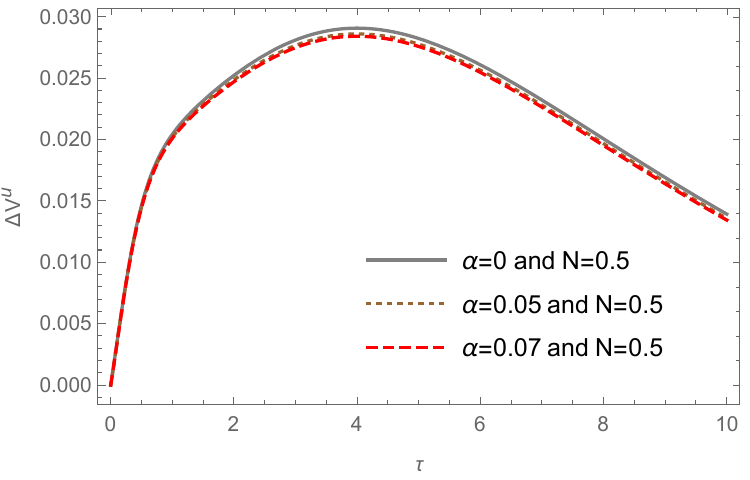}
    \caption{  The variation of $\Delta V^{u}$ in the presence of a GW pulse for different values of $\alpha$ in the case $\omega=0$ with surrounding field parameter $N=0.5$.}
  \end{minipage}
\end{figure}

Then, the resulting geometry which is the sum of black hole geometry and GW perturbation becomes
\begin{equation}
    ds^{2}=-f(r)du^{2}-2dudr +(r^{2}+rH(u))d\theta^{2}+(r^{2}-rH(u))\sin^{2}\theta d\phi^{2},
\end{equation}
where
\begin{equation}
    f(r)=1-\frac{2m}{r}+\frac{N}{r^{3\omega+1}}.
\end{equation}

Note that  $H(u)$ corresponds to GW pulse and is taken to be
\begin{equation}
    H(u)=A sech^{2}(u-u_{0}),
\end{equation}
where $A$ indicates the amplitude of the GW pulse which is centered around $u=u_{0}$.
The GW pulse is represented by the function $H(u)$, while the metric $h_{\mu\nu}$ is assumed to be expressed in the transverse-traceless $(TT)$ gauge. To introduce this perturbation into our metric, we must incorporate an energy momentum tensor in the matter sector, which serves as the source for the GW pulse $H(u)$ in the $TT$ gauge. This energy momentum tensor can be derived from an expanding anisotropic fluid shell. Prior to the perturbation, the fluid was not dynamic, and the hypersurfaces with constant $u$ were spherically symmetric. However, as a result of this expansion, the GW pulse is generated and propagates throughout the spacetime towards future null infinity.

On the equatorial plane $\theta =\frac{\pi}{2}$, the geodesic equations for the $u$ coordinate becomes, 
\begin{equation}\label{geo1}
    \ddot{u}-\frac{f^{\prime}}{2}\dot{u}^{2}-\frac{H-2r}{2}\dot{\phi}^{2}=0
\end{equation}
For $r$ and $\phi$ coordinates we have following geodesic equations
\begin{equation}\label{geo2}
    \ddot{r}+\frac{f}{2}f^{\prime}\dot{u}^{2}+f^{\prime}\dot{r}\dot{u}+\frac{fH-2fr-rH^{\prime}}{2}\dot{\phi}^{2}=0
\end{equation}

\begin{equation}\label{geo3}
   \ddot{\phi}-\frac{H^{\prime}}{r-H} \dot{\phi}\dot{u}+\frac{2r-H}{r(r-H)}\dot{\phi}\dot{r}=0
\end{equation}
where $H^{\prime}=\frac{dH}{du}$ and $f^{\prime}=\frac{df}{dr}$.

The background geometry which is the hairy Kiselev black hole considering for large $r$, on the equatorial plane $\theta=\frac{\pi}{2}$ satisfies the following condition  

\begin{equation}\label{geo4}
    -f(r) \dot{u}^{2}-2\dot{u}\dot{r}+r^{2}\dot{\phi}^{2}-rH(u)\dot{\phi}^{2}=-1
\end{equation}
Numerical methods are required to solve the non-linear equations $(\ref{geo1})$,$(\ref{geo2})$ and $(\ref{geo3})$ in order to determine two neighboring geodesics in the presence of a GW pulse. The initial conditions for the two geodesics are selected as follows: the initial values of $r$ and $u$ for geodesics I and II are arbitrary, but $\phi$ is the same for both. Additionally, the initial conditions for  $\dot{r}$ and $\dot{u}$ are fixed for the geodesics. As we are limited to the equatorial plane, the value of  $\dot{\phi}$  can be determined using equation  $(\ref{geo4})$.

The calculation of the difference between two geodesics is given by the equations:
\begin{equation}
  \Delta r=r(geodesic II)-r(geodegic I) ~~~,~~~ \Delta u=u(geodesic II)-u(geodesic I).  
\end{equation}
In the case of a dust-like field for Kiselev solution where the constant equation of the state parameter of
the surrounding field is $\omega=0$, the values of $\Delta r$ and $\Delta u$ are shown in figures $(3)$ and $(4)$ respectively. It can be observed that as the parameter $N$ increases, the deviation between the neighboring coordinates of $r$ decreases. On the other hand, for the $u$ coordinate, as $N$ increases, the value of $\Delta u$ increases. This indicates that $\Delta r$ and $\Delta u$ exhibit opposite behaviors with respect to $N$. 
To illustrate figures  $(3)$ and $(4)$, the pulse profile is selected with $u_{0}=0$ and $A_{0}=1$, while the mass $m$ is assigned a value of $1$.

 In order to obtain a black hole as the central compact object, as has been mentioned in this section, we impose a restriction on the constant equation of the state parameter of the surrounding field, denoted as $\omega > -\frac{1}{3}$. This condition ensures that the solution we are interested in is asymptotically flat. Another scenario we consider involves the surrounding field being radiation, characterized by $\omega =\frac{1}{3}$. In this case, the Kiselev black hole solution is further constrained by the condition $m^{2}\ge |N|$, which guarantees the existence of a horizon \cite{yagoub}. Furthermore, we can explore the possibility of a surrounding field with a constant parameter  $\omega = \frac{2}{3}$. In this scenario, the falling order of this field is greater than that of the radiation field at null infinity or for large values of $r$.

For the parameters $\omega=\frac{1}{3}$ and $\omega=\frac{2}{3}$, the corresponding values of $\Delta r$  and $\Delta u$  can be found in figures $(5)$ to $(8)$. By analyzing these figures, it can be concluded that the detection of memory effects of the GW pulse becomes more challenging as the parameters $\omega$ increase, especially for various values of the surrounding field parameter $N$. As depicted in the figures, the change in $\Delta u$  is more prominent compared to $\Delta r$. In the range of figures $(5)$ to $(8)$, the initial values are set as  $A_{0}=1$, $u_{0}=0$ and $m=2$. It is worth noting that, upon observing figures $(3)$ to $(8)$, it becomes apparent that the variations in  $\Delta r$  and $\Delta u$ for different values of the surrounding field parameter $N$ are not easily distinguishable as $\omega$ increases. This phenomenon can be attributed to the rapid decline of the term with a large value of $\omega$ in the asymptotic region.  Moving forward, we will explore the emergence of displacement and velocity memory effects in the Kiselev background geometry.

The presence of the displacement memory effect can be observed by examining figures $(9)$ to $(14)$, which clearly demonstrate that the disparities $\Delta u$ and $\Delta r$ between adjacent timelike geodesics are influenced by the passage of a GW pulse. Specifically, the values of  $\Delta u$ and $\Delta r$ do not revert back to those of the black hole background after the GW has traversed, thus establishing the occurrence of the displacement memory effect. The underlying spacetime geometry has a strong influence on this effect, which is not merely determined by the intensity of the GW pulse. Varying the parameter $N$ will result in distinct deviations and different manifestations of the memory effect.

The memory effect holds potential as a viable means to examine the presence of hairs of a black hole or any compact object. Additionally, the influence of the coupling parameter $\alpha$ and the length parameter $l$, which are encoded in the mass as $m=\mathcal{M}+\frac{\alpha l}{2}$, can be taken into consideration to ascertain the changes in $\Delta u$ and $\Delta r$ as displacement memory effects in  $\Delta V^{r}$ and $\Delta V^{u}$ as velocity memory effects. These effects are illustrated in figures $(27)$ to $(30)$. In the current context, it signifies that non-zero values of $l$ or $\alpha$ do indeed impact the displacement and velocity memory in neighboring geodesics. As depicted in figures  $(27)$ to $(30)$, the change in $\Delta u$ is more pronounced in comparison to $\Delta r$. For dust-like surrounding field with parameter $\omega=0$, the figures are illustrated with the initial conditions as $u_{0}=0$, $A_{0}=1$, $l=0.5$ and $\mathcal{M}=0.5$. 

In addition to the displacement memory effect, the Kiselev black hole geometry exhibits a velocity memory effect, as evident from the depiction in Figures $(21)$ to $(26)$. Both $\Delta V^{r}$ and $\Delta V^{u}$ exhibit non-zero values and deviate from their background values when influenced by the GW pulse. This memory effect also depends on the choice of the specific surrounding fields parameter $N$ and the different constant equations of the state parameter of the surrounding field $\omega$.

It is evident from the data presented in figures $(15)$ to $(20)$ that the velocity memory effects, denoted by   $\Delta V^{r}$, decrease as the value of $N$ increases. Conversely, the variation of  $\Delta V^{u}$ increases with an increase in $N$. Furthermore, the velocity memory effects are more pronounced for the $\dot{u}$ component and become less discernible as the surrounding field parameter $\omega$ increases. 

The investigation of the surrounding field parameter, in addition to the potential inclusion of a comprehensive examination of the displacement and velocity memory effect, has the potential to establish an undeniable confirmation of non-zero values of parameters associated with the hair of the hairy Kiselev black hole solution. Furthermore, it can provide insight into the surrounding field parameter $N$ and the constant of the equation of the state parameter $\omega$. As a result, it can be inferred that both the displacement and velocity memory effects exist within the context of the hairy Kiselev black hole solution, and their existence is intricately connected to the selection of $N$, $\omega$, $l$, and $\alpha$. This discovery offers a new avenue for exploring the hair of compact objects, particularly black holes, which in turn may yield valuable insights into the potential existence of physics beyond the realm of general relativity. 
  

\section{Conclusion}

The Kiselev black hole solution, characterized by its hairy Schwarzschild-like appearance and surrounded by a field with field parameter $N$, presents an intriguing possibility for exploring theories of gravity and spacetime beyond classical general relativity. This solution challenges the No-Hair theorem of general relativity, as its unknown source field may violate the theorem and suggest new fundamental fields. To study the memory effects of gravitational waves, the Kiselev black hole solution is considered as a background that is perturbed by a gravitational wave pulse at null infinity and for large values of $r$. This is achieved by studying the memory effect of the Kiselev black hole at null infinity using Bondi-Sachs formalism. The gravitational wave pulse is considered in the background of the hairy Kiselev solution to observe the change in Bondi mass, which is a measure of mass on an asymptotically null slice or the densities of energy on celestial spheres, for a certain value of the constant of the equation of state parameter $\omega=0$. The field parameters and hairy Kiselev black hole parameters have effects on the change of Bondi mass, and the changes in Bondi mass at null infinity are indicated for particular values of $N$, the constant equation of the state parameter of the surrounding field $\omega$, and parameters related to the hair of the hairy Kiselev solution, such as the coupling constant $\alpha$ and length parameter $l$, both in the presence and absence of the gravitational wave pulse. These results are presented in figures $(1)$ and $(2)$. As the length parameter $l$ grows, the magnitude of the variation in the change of Bondi mass becomes notable as illustrated in figure $(2)$. This variation is observed at future null infinity and for substantial $r$, indicating the impact of the black hole's hair at a considerable distance from it.  Therefore, the hairy black hole spacetime serves as a phenomenological model for testing the hypothesis of new fundamental fields, and we explore how the coupling constant $\alpha$ and the length parameter $l$ affect the Bondi mass in our model of interest.  The figure $(1)$ illustrates the impact of the background field on the changes in the Bondi mass. It is evident that as the field parameter $N$ increases, the magnitude of the change in Bondi mass decreases. This observation aligns with the fact that $N$ is a positive value and its influence opposes the behavior of the mass $m$. In simpler terms, the gravitational attraction, which is governed by $m$, is weakened by the increase in $N$. 

By examining the displacement and velocity memory effects, as well as the surrounding field parameters, it is possible to confirm the non-zero values of parameters associated with the hair of the hairy Kiselev black hole solution. In our study, we investigated the deviation of neighboring geodesics for the hairy Kiselev black hole at a large distance from it, with consideration given to the surrounding field parameter $N$, the constant equation of the state parameter of the surrounding field $\omega$, and parameters related to the hair of the solution, such as the coupling constant $\alpha$ and length parameter $l$. Our results, which were numerically solved and depicted in figures $(3)$ to $(30)$, indicate that these parameters significantly impact the deviation of neighboring geodesics, particularly at the boundary of a compact object such as a black hole. With the increase of the surrounding field parameter $N$, the displacement memory effects $\Delta r$ decrease, whereas the displacement $\Delta u$ increases. When a GW pulse is present in the background, the change in displacement memory effect $\Delta r$ increases, but the opposite effect is observed for $\Delta u$. The velocity memory effects $\Delta V^{r}$ and $\Delta V^{u}$ exhibit similar behavior to $\Delta r$ and $\Delta u$, respectively.  As the coupling constant $\alpha$ increases, the displacement memory effects $\Delta r$ show an increase, while the displacement $\Delta u$ experiences a decrease. Similarly the velocity memory effects $\Delta V^{r}$ and $\Delta V^{u}$ demonstrate a similar behavior to $\Delta r$ and $\Delta u$, respectively. This discovery not only confirms the existence of displacement and velocity memory effects within the context of the hairy Kiselev black hole solution but also provides valuable insights into the potential existence of physics beyond the realm of general relativity.

  \section*{Acknowledgements}

This research is supported by the research grant of the University of Tabriz (sad/779).
DFM thanks the Research Council of Norway for their support, and UNINETT Sigma2 -- the National Infrastructure for High Performance Computing and
Data Storage in Norway.

\section*{Data availability statement}
No new data were created or analyzed in this study.


\end{document}